\documentclass[%
reprint,
twocolumn,
 amsmath,amssymb,
 aps,
]{revtex4-2}

\usepackage{graphicx}
\usepackage{dcolumn}
\usepackage{bm}
\usepackage{layouts}
\usepackage{float}
\usepackage{comment}
\usepackage{soul}

\usepackage[dvipdfm,colorlinks,linkcolor=blue,anchorcolor=blue,citecolor=blue,urlcolor=blue]{hyperref}

\begin{document}

\preprint{APS/123-QED}

\title{Measurement of the Electric Dipole Moment of $^{171}$Yb Atoms in an Optical Dipole Trap}

\author{T. A. Zheng$^{1}$}
\thanks{These authors contributed equally to this work.}
\author{Y. A. Yang$^{1}$}%
\thanks{These authors contributed equally to this work.}
\author{S.-Z. Wang$^{1}$}
\author{J. T. Singh$^{2}$}
\author{Z.-X. Xiong$^{3}$}%
\author{T. Xia$^{1}$}%
\email{txia1@ustc.edu.cn}
\author{Z.-T. Lu$^{1}$}%
\email{ztlu@ustc.edu.cn}

\affiliation{$^1$CAS Center for Excellence in Quantum Information and Quantum Physics, University of Science and Technology of China, Hefei 230026, China}%
\affiliation{$^2$National Superconducting Cyclotron Laboratory and Department of Physics and Astronomy,
Michigan State University, East Lansing, Michigan 48824, USA}%
\affiliation{$^3$Key Laboratory of Atomic Frequency Standards, Innovation Academy for Precision Measurement Science and Technology, Chinese Academy of Sciences, Wuhan 430071, China}


\date{\today}

\begin{abstract}
The permanent electric dipole moment (EDM) of the $^{171}$Yb $(I=1/2)$ atom is measured with atoms held in an optical dipole trap (ODT). By enabling a cycling transition that is simultaneously spin-selective and spin-preserving, a quantum non-demolition measurement with a spin-detection efficiency of 50$\%$ is realized. A systematic effect due to parity mixing induced by a static E field is observed, and is suppressed by averaging between measurements with ODTs in opposite directions. The coherent spin precession time is found to be much longer than {\color{black}300 s}. The EDM is determined to be $d({\rm^{171}Yb})={\color{black}(-6.8\pm5.1_{\rm stat}\pm1.2_{\rm syst})\times10^{-27}\ e\ \rm cm}$, leading to an upper limit of $|d({\rm^{171}Yb})|<{\color{black}1.5\times10^{-26}\ e\ \rm cm}$ ($95\%$ C.L.). These measurement techniques can be adapted to search for the EDM of $^{225}$Ra.

\end{abstract}

\maketitle

$\textit{Introduction\ -\ }$The existence of a permanent electric dipole moment (EDM) of an atom or a subatomic particle violates the time-reversal symmetry ($T$) \cite{ChuFieRam19,PurRam50,IosSte97} and, under the $CPT$ theorem, the $CP$ symmetry as well \cite{ChrCroFit64,AubBouGai01,AbeAbeAda01}. $CP$-violation sources in the Standard Model contribute to EDMs only at higher orders, resulting in EDM predictions far below the experimental reach in the foreseeable future \cite{IosSte97}. On the other hand, Beyond-Standard-Model (BSM) scenarios, such as Supersymmetry, naturally provide additional sources of $CP$ violation that can potentially induce EDMs large enough to be observed in the current generation of experiments \cite{ForSanBar03}. 

{\color{black}EDM searches primarily belong to three categories: EDM of the electron, EDM of the nucleons, and nuclear Schiff moments. Recent measurements have placed upper limits on the electron EDM \cite{AndAngDeM18,CaiGreGra17,HudKarSma11}, the neutron EDM \cite{AbeAfaAyr20,BakDoyGel06}, and the EDM of the diamagnetic atom $^{199}$Hg \cite{GraCheLin16,GraCheLin17}, leading to tight constraints on BSM $CP$-violating interactions \cite{JonMic13,ChuRam15}. The three categories are complementary to each other as they are sensitive to different sources of new physics \cite{ChuFieRam19}.} EDMs in diamagnetic atoms and molecules are primarily sensitive to $CP$-violating interactions that induce nuclear Schiff moments \cite{MaxAda05,Sch63,GinFla04}. Within this category, the EDM of $^{199}$Hg is measured in a vapor cell \cite{GraCheLin16}, $^{129}$Xe in a gas cell \cite{SacFanBab19}), $^{205}$Tl$^{19}$F in a molecular beam \cite{ChoSanHin91}, and the EDM of laser-cooled $^{225}$Ra atoms are probed in an optical dipole trap (ODT) \cite{ParDieKal15,BisParBai17}. Among them, the best limits on new physics are derived from the atomic EDM of $^{199}$Hg: $|d({\rm ^{199}Hg})|<7\times10^{-30}\ e\ \rm cm$ \cite{GraCheLin16}. EDMs of the other systems, although at lower levels of precision, are combined with the $^{199}$Hg result to provide tighter constraints on multiple hadronic $CP$-violating parameters \cite{ChuRam15}. Due to its nuclear octupole deformation \cite{AueFlaSpe96,GafButSch13}, $^{225}$Ra is an attractive case as its EDM is predicted to be three orders of magnitude larger than that of $^{199}$Hg \cite{DobEng05}. However, its radioactivity and rarity cause considerable difficulties in the development of the cold-atom ODT method for the EDM measurement. The current limit is $|d({\rm ^{225}Ra})|<1.4\times10^{-23}\ e\ \rm cm$ \cite{BisParBai17}.

{\color{black}Measuring EDM on laser-cooled atoms in an ODT has important advantages. On one hand, the EDM sensitivity is boosted by a high electric field and a long spin precession time; on the other, the $\bf{v\times E}$ systematic in a beam and the leakage-current systematic in a cell are both avoided. A detailed analysis of the ODT method concluded that the measurement precision had the potential of reaching $10^{-30}\ e\ \rm cm$ \cite{RomFor99}. The stable and abundant $^{171}$Yb is among the few diamagnetic isotopes having nuclear spin $I=1/2$, thus eliminating tensor effects. Its atomic structure is ideally suited for the development of the cold-atom ODT method that probes for the type of EDM that originates from the nuclear Schiff moment.} Indeed, EDM measurements of $^{171}$Yb in an atomic fountain \cite{Nat05} or an ODT \cite{Tak97} were both proposed, but no measurements had previously been carried out. In this letter, we present the first experimental limit on the atomic EDM of $^{171}$Yb. By enabling a spin-selective and spin-preserving cycling transition, a quantum non-demolition measurement is realized to improve the spin-detection efficiency by a factor of 50. A systematic effect due to parity mixing induced by a static E field \cite{RomFor99} is observed for the first time, and is suppressed by averaging between measurements with ODTs propagating in opposite directions. {\color{black}The decoherence rate of spin precession is reduced to the level of $1\times 10^{-4}$ s$^{-1}$. As shown in Table~\ref{tab:table2}, the contribution of this first $^{171}$Yb EDM limit to constraining BSM physics, although lagging behind that of $^{199}$Hg, is on the same order of magnitude as those of $^{129}$Xe, $^{225}$Ra, and $^{205}$Tl$^{19}$F.}

\begin{figure}[t]
	\includegraphics[width=3.4in]{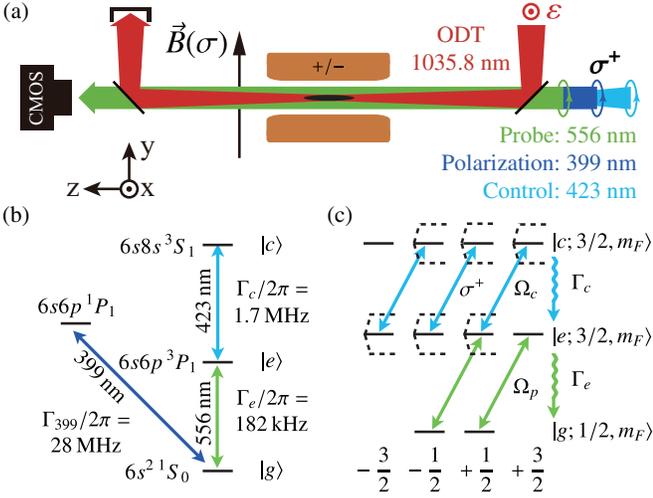}
	\caption {(a) Layout of the setup. (b) Energy levels and transitions of  Yb. (c) The QND approach for efficient spin detection. With the control laser beam dressing the $|e\rangle\leftrightarrow|c\rangle$ transitions, the state $|g; 1/2, -1/2\rangle$ becomes a dark state, while $|g; 1/2, +1/2\rangle$ can be probed repeatedly on a cycling transition. Each spin state is preserved and probed without risking spin flips.}
\label{Appar_and_tran}
\end{figure}

$\textit{Experimental\ setup\ -\ }$In the experiment, neutral $^{171}$Yb atoms are loaded into a two-stage magneto-optical trap (MOT), and laser cooled on the narrow-line transition $6s6p\ ^{1}S_0\leftrightarrow6s6p\ ^{3}P_1$ to 20 $\mu$K \cite{ZheYanSaf20}. The atoms are then transferred to a moveable ODT that carries them over a distance of 65 cm into a neighboring science chamber \cite{ParDieBai12}, and are handed over to a stationary ODT (power {\color{black}10 W}, waist 50 $\mu$m, Rayleigh length 6.1 mm, trap depth {\color{black}60 $\mu$K}). The moveable ODT is then turned off and EDM measurements are performed with atoms held in the stationary ODT [Fig.~\ref{Appar_and_tran} (a)]. Under 10$^{-11}$ Torr in the science chamber, the atoms have a trap lifetime of {\color{black}75 s}. The two ODTs are provided by two separate fiber laser amplifiers, both at the magic wavelength of 1035.8 nm for the $6s6p\ ^{1}S_0\leftrightarrow6s6p\ ^{3}P_1$ transition \cite{ZheYanSaf20}, thus suppressing both the shift and broadening of the transition used for spin-state detection.

As shown in Fig.~\ref{Appar_and_tran} (a), the stationary ODT is linearly-polarized in the $x$ direction, and propagates in the $z$ direction through the {\color{black}1.82(5) mm} gap between a pair of parallel copper electrodes whose circular end faces are 1.6 cm in diameter. The lower electrode is grounded; the upper one can be ramped to voltages between {\color{black}$+$13.3 kV} and {\color{black}$-$13.3 kV}, generating a uniform E field of {\color{black}$\pm$73 kV/cm} in the $y$ direction. The leakage current measured on the grounded side is typically $<$ {\color{black}2 pA}. The electrode assembly is held inside a titanium UHV chamber, which in turn is surrounded by a $\cos\theta$ coil. $\mu$-metal shields provide a $y$-direction shielding factor of $4\times10^4$ at the center. The coils inside the shields generate a stable and uniform B field of 20 mG in the $y$ direction. Its spatial nonuniformity is less than $5\times 10^{-4}$ cm$^{-1}$, and temporal instability less than 1 ppm when averaged over the load-measurement cycle of about {\color{black}125 s}.

$\textit{QND\ measurement\ of\ spin\ precession\ -\ }$The Larmor precession frequencies ($f_{\pm}$) of the atoms are measured: $hf_{\pm}=2\mu B\pm 2dE$, where $\mu$ is the magnetic dipole moment ({\color{black}$+$0.49367(1) $\mu_N$} for $^{171}$Yb \cite{Table05}), $d$ is the EDM, and $f_{+}$($f_{-}$) is the precession frequency corresponding to the case of the E field being parallel (antiparallel) to the B field. The absorption of a resonant probe beam is used to determine the precession phase (Fig.~\ref{Appar_and_tran}). It differentiates the $m_F=\pm1/2$ states (quantization axis chosen to be in the $z$ direction) in the ground level by turning one into a “dark state” that does not absorb photons, and the other into a “bright state”. Usually, as in the $^{225}$Ra EDM experiment \cite{ParDieKal15}, the bright state only undergoes on average a few excitation-decay cycles before a spin flip occurs due to optical pumping. This effect severely limits the spin-state detection efficiency. We propose and demonstrate a quantum non-demolition (QND) scheme for spin-state detection in order to achieve a much higher detection efficiency. As shown in Fig.~\ref{Appar_and_tran} (b) and \ref{Appar_and_tran} (c), the three levels $6s^2\ {}^1S_0\ (F=1/2)$, $6s6p\ {}^3P_1\ (F=3/2)$ and $6s8s\ {}^3S_1\ (F=3/2)$ form a $|g\rangle-|e\rangle-|c\rangle$ ladder system. A $\sigma^{+}$ polarized “control” laser beam is introduced to resonantly dress the $|e\rangle\leftrightarrow|c\rangle$ transition with the Rabi frequency $\Omega_{c}$. Among the Zeeman states $|e;\,3/2,\,m_F\rangle$, only the stretched state $|e;\,3/2,\,+3/2\rangle$ is unaffected; all the other states are dressed by the control beam and their corresponding energy levels shifted by $\pm\ \Omega_{c}/2$ to form Aulter-Townes doublets~\citep{KhaBhaNat16}. Meanwhile, a $\sigma^{+}$ polarized “probe” laser beam is introduced to resonantly drive the $|g\rangle\leftrightarrow|e\rangle$ transition with the Rabi frequency $\Omega_{p}$. In case of a weak probe, i.e. $\Omega_{p}\ll\Omega_{c}\ {\rm or}\ \Gamma_{e}$, the absorption rate for $|g; 1/2, -1/2\rangle$ is reduced by a factor of $\sim \Omega_{c}^{2} / (\Gamma_{e}\Gamma_{c})$, and the spin flip rate is suppressed  by the same factor. In such an arrangement, $|g;\,1/2,\,-1/2\rangle$ remains a dark state, while $|g;\,1/2,\,+1/2\rangle$ can be excited repeatedly via a near-cycling transition without inducing a spin flip. Thus, a QND measurement is realized.

The laser beams at 556 nm and 399 nm are supplied by two frequency-doubled diode lasers and the control beam at 423 nm is supplied by a frequency-doubled Ti:Sapphire laser. The probe, control and polarization beams all have the same $\sigma^{+}$ circular polarization, and all co-propagate along the $z$ direction. The control beam is focused onto the atoms with a beam waist of 300 $\mu$m, and the parameters for the control beam are determined by measuring the resulting light shifts of the probe transition. At the control beam power of 40 mW ($\Omega_{c}\sim 2\pi\times40$ MHz), the spin-flip rate in the ground level is reduced by a factor of $\Omega_{c}^{2}/(\Gamma_{e}\Gamma_{c})\sim 10^{3}$. This QND scheme improves the spin state detection efficiency to 50$\%$, which is 50 times higher than the conventional non-QND approach.

\begin{figure}[t]
	\includegraphics[width=3.4in]{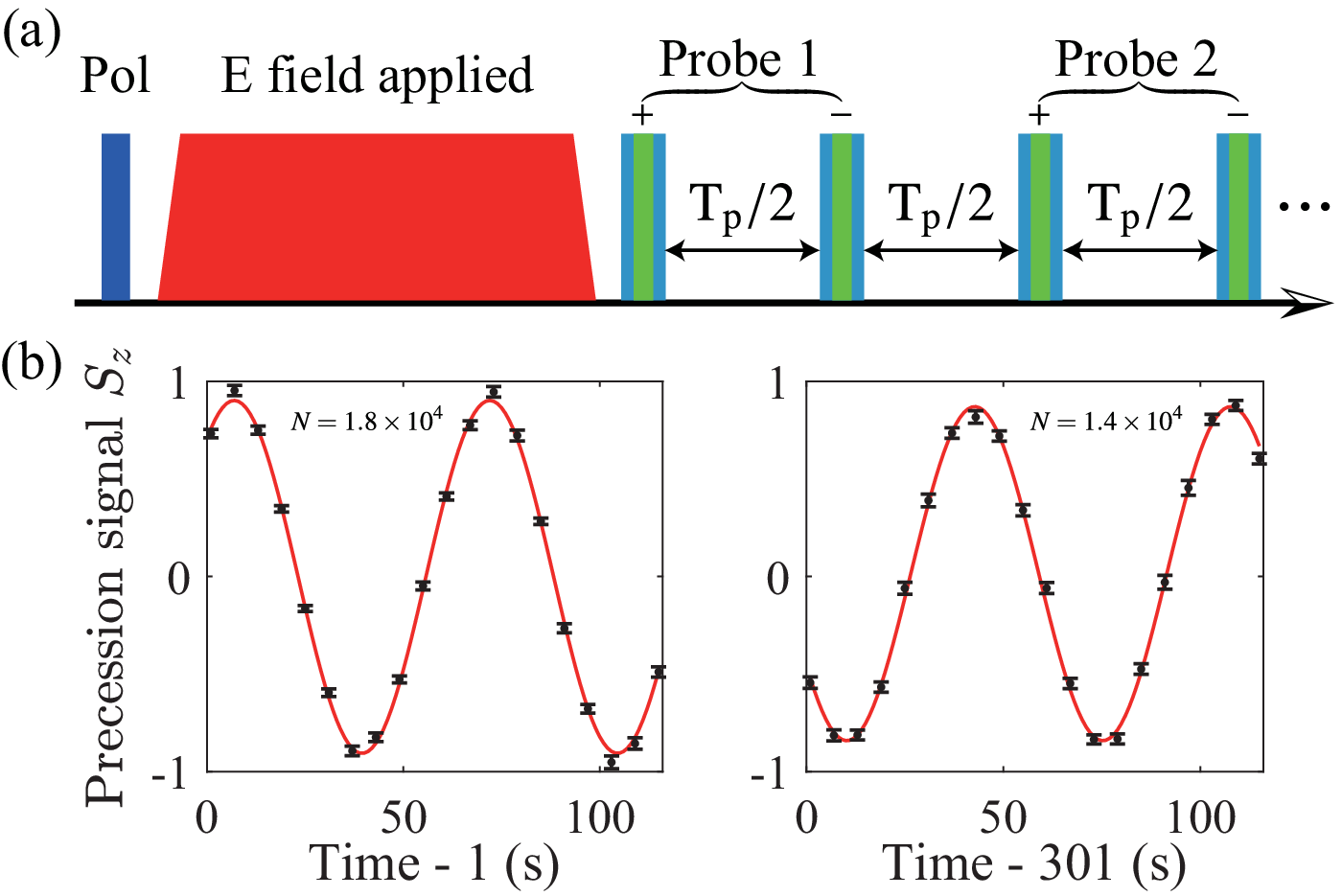}
	\caption {(a) Timing sequence of the spin precession measurement. Atoms are initially polarized by a 399 nm pulse (“Pol”, $\color[rgb]{0.11,0.33,0.67}\blacksquare$). After a given precession time, two 556 nm ($\color[rgb]{0.39,0.768,0.298}\blacksquare$) $+$ 423 nm ($\color[rgb]{0.204,0.612,0.8}\blacksquare$) probe pulses (“Probe 1”), separated by $T_{p}/2$, measure the populations in $m_{F}=\pm1/2$ states successively. The E field ($\color[rgb]{1,0.20,0.11}\blacksquare$) is applied between the polarization and probing.
	(b) The precession signals at 1 s and {\color{black}301 s}. The precession frequency is 15 Hz at a bias \affiliation{} field of 20 mG.} 
\label{Sequ_Cohe}
\end{figure}

$\textit{EDM\ measurements\ -\ }$The pulse sequence for the EDM measurement is shown in Fig.~\ref{Sequ_Cohe} (a). The spin polarized ensemble is initially prepared with a 2 ms long pulse of the polarization beam (“Pol”, $I/I_{s}=3\times10^{-4}$, {\color{black}where $I$ is the laser intensity and $I_{s}$ is the saturation intensity of the transition}) resonant with the non-cycling transition of $6s6p\ ^{1}S_0,F=1/2\leftrightarrow6s6p\ ^{1}P_1,F=1/2$. The atoms precess about the bias B field at a Larmor frequency of $\sim$ $2\pi\times$15 Hz. After a given precession time, a 0.4 ms long overlapping pulse of the probe beam ($I/I_{s}={\color{black}0.03}$) and the control beam ($I/I_{s}=5\times10^{3}$), named “Probe 1$+$”, is applied for a spin-selective absorption measurement. The population in $|g;\,1/2,\,+1/2\rangle$ ($N_{+}$) is measured, while the population in $|g;\,1/2,\,-1/2\rangle$ ($N_{-}$) remains in the dark. Half of a period ($T_{p}/2$) later, the precession swaps $N_{+}$ and $N_{-}$, and an identical probe pulse (Probe 1$-$) is fired to measure the original $N_{-}$ prior to swapping. Since the populations are preserved during the QND measurement, the probe pulses can be repeated, each with a $T_{p}/2$ delay from the previous pulse. {\color{black}As a result, the signal-to-noise ratio is greatly enhanced such that the measurement is limited by the quantum projection noise.} Following a total of {\color{black}8} “Probes” ({\color{black}16} pulses), the atoms in the ODT are dropped and 30 background images are taken for fringe-removal \cite{OckTauSpr10}. 

The precession signal is expressed as 
\begin{equation}
S_z=\frac{N_+-N_-}{N_++N_-}=Ce^{-\frac{t}{T_2}}\cos(2\pi ft+\phi_0)+O,\label{deltanu}
\end{equation}
where $C=0.90(1)$ is the precession amplitude at 1 s [Fig.~\ref{Sequ_Cohe} (b)], $T_2$ is the spin coherence time, $f$ is the precession frequency, $\phi_0$ is the initial phase and $O$ the offset. As shown in Fig.~\ref{Sequ_Cohe} (b), the precession amplitude is reduced to {\color{black}0.86(1)} after {\color{black}300 s}, indicating that the spin
coherence time ($T_2$) is much longer than {\color{black}300 s}, and is estimated to be {\color{black}$(9\pm4)\times10^3$ s}. In this work, the overall interrogation time is limited by the trap lifetime of {\color{black}75 s}.

\begin{figure}[b]
	\includegraphics[width=3.4in]{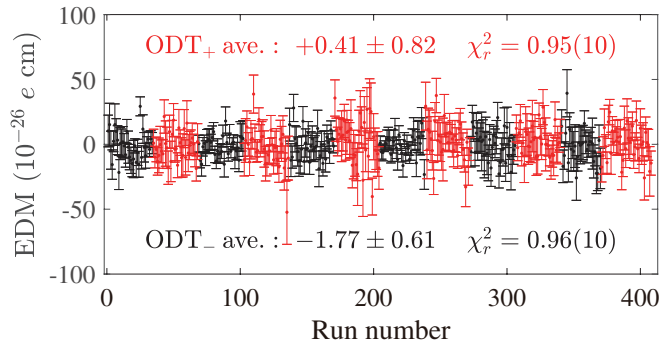}
	\caption {EDM results. {\color{black}The data are presented in order of data taking, with each point taking 80 minutes, corresponding to a proper bin size of 18 \cite{Hutzler19} . The red data points and analysis results are for measurements made with ODT$_+$, and the black ones for ODT$_-$. The final EDM result is the average of both the ODT$_+$ and ODT$_-$ data.}}
\label{EDM_stat_all}
\end{figure}

In each E field polarity, following a precession over time $\tau$, the phase advance $\phi^+(\phi^-)$ is measured. Combining the two measurements, the phase shift correlated with the E field reversal,
$\Delta\phi=\phi^+-\phi^-$, is used to calculate the EDM and its systematics $d={\hbar\Delta\phi}/(4E\tau)$, where $E$ is the static E field. The precession phases are probed at the optimum points when $S_z(t)=0$. The E field polarity is randomly switched between ($+-$) and ($-+$) patterns for any two sequential cycles. Every few days, the measurement point is switched between positive and negative $S_z$ slope. {\color{black}In order to mitigate the residual parity-mixing systematic effect, we perform the measurements in two ODTs with opposite $k$ vectors: ODT$_+$ (traveling in the $+z$ direction in Fig.\ref{Appar_and_tran} (a)) and ODT$_-$, and switch between them every other day. EDM data are accumulated for {\color{black}a total of 260 h} for ODT$_+$ and {\color{black}250 h} for ODT$_-$ (Fig.\ref{EDM_stat_all}).} The statistical uncertainty can be expressed as
\begin{equation}
\delta d=\frac{\hbar}{2E\tau\sqrt{ n}}\sqrt{\frac{1}{N\epsilon}+\sigma^2_{\phi(\delta B)}}.
\label{EDM_stat}
\end{equation}
The precession time $\tau_{+}=\tau_{-}={\rm {\color{black}96\ s}}$, the number of repeated measurements $n_{+}={\rm \sim{\color{black}7,500}}$ and $n_{-}={\rm \sim{\color{black}7,200}}$, the atom number $N_+={\color{black}5.9(1.6)\times10^4}$ and  $N_-={\color{black}4.9(1.8)\times10^4}$, and the spin-detection efficiency $\epsilon\approx\rm{0.5}$. Over the precession time of 96 s, the measured phase noise $\sigma_{\phi(\delta B)}$ is induced by B field noise of $\sim$ 3 pT. Its contribution to the EDM uncertainty is on par with the $1/\sqrt{N\epsilon}$ term. {\color{black}The B field noise is primarily caused by the Seebeck effect due to heating of the titanium chamber and the electrodes by both the moveable and the stationary ODTs}.

\begin{figure}[b]
	\includegraphics[width=3.4in]{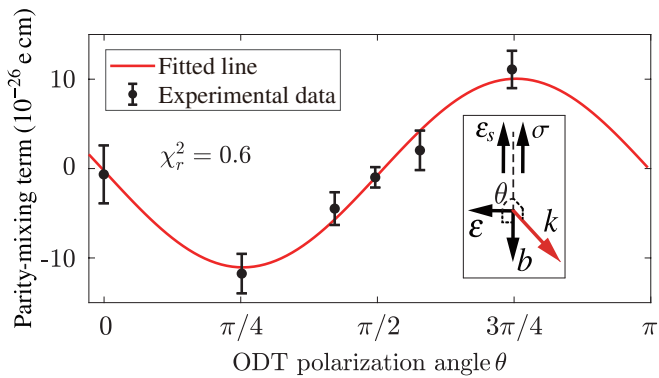}
	\caption {The parity-mixing  systematic measured at different ODT polarization angles. $k$ denotes the propagation direction of the ODT.}
\label{E1M1_measurement}
\end{figure}

$\textit{Parity\,-\,mixing\ effect\ -\ }$The EDM measurement in an ODT is predicted to experience a unique systematic effect due to parity mixing induced by the strong static E field \cite{RomFor99, BisParBai17}. It originates from the coupling between the optical fields of the ODT and the static E field, resulting in a vector shift ($\Delta \nu$) in the ground state that is linearly proportional to both the E field and the ODT intensity \cite{RomFor99}:
\begin{equation}
\Delta\nu=\nu_1(\hat{b}\cdot\hat{\sigma})
(\hat{\varepsilon}\cdot\hat{\varepsilon_s})+
\nu_2(\hat{b}\cdot\hat{\varepsilon_s})
(\hat{\varepsilon}\cdot\hat{\sigma}),\label{deltanu}
\end{equation}
where $\hat{b}$ and $\hat{\varepsilon}$ denote the directions of the ODT magnetic and electric fields, respectively. $\hat{\sigma}$ is the spin quantization axis direction defined by the bias B field, and $\hat{\varepsilon_s}$ is the direction of the static E field. The $\nu_1$ and $\nu_2$ coefficients depend on the atomic properties, the E field and the ODT intensity \cite{RomFor99}. The directions of the various fields in our experiment [Fig.~\ref{Appar_and_tran} (a)] are designed to zero the parity-mixing systematic, yet imperfection in field alignment causes a residue effect.

Here we aim to reveal this effect and examine its dependence by varying the angle $\theta$ between the static fields and the ODT polarization vector (Fig.~\ref{E1M1_measurement}). Assuming $\hat{\sigma}\cdot\hat{\varepsilon_s}=1$ and $\hat{\varepsilon_s}\cdot\hat{k}=0$, Eq. (\ref{deltanu}) can be simplified to
\begin{equation}
\Delta\nu=-\frac{\nu_1+\nu_2}{2}\sin(2\theta),\ 0\leq\theta<\pi. \label{deltanu_theta}
\end{equation}
By fitting the data, we determine the maximum vector shift expressed in terms of a false EDM, $d_{false}=1.06(0.17)\times10^{-25}\,e\,\rm{cm}$ (Fig.~\ref{E1M1_measurement}). This value is on the same order of magnitude as the result of a partial calculation on $^{199}$Hg in Ref \cite{RomFor99}.


Experimentally, it is difficult to determine all the field directions in order to set the angle $\theta$ at exactly $\pi/2$. Instead, as described in the previous section, two counter-propagating ODTs, ODT$_+$ and ODT$_-$, are employed to suppress the systematic effect. EDM measurements are performed with a common polarization angle, both set to be as close as possible to the ideal condition of $\theta_+=\theta_-=\pi/2$ (Fig.~\ref{E1M1_measurement}). The difference $\theta_+-\theta_-$ can be determined using a common polarizer to an accuracy of 7 mrad, while each angle itself can only be determined to an accuracy of 80 mrad. Upon $k$ reversal, the parity-mixing systematic effect reverses sign along with the $\hat{b}$ vector, while the true EDM remains the same. Therefore, the residual parity-mixing systematic effect is suppressed by averaging the EDM results in ODT$_+$ and ODT$_-$.

\begin{table}[b]
\caption{\label{tab:table1}%
Systematic effects on the measured EDM. All entries are specified in the unit of ($10^{-27}e\ {\rm cm}$).}
\begin{ruledtabular}
\begin{tabular}{lc}
Contribution&Uncertainty\\
Bias B-field correlation &{\color{black}1.00}\\
Residual parity-mixing effect &{\color{black}0.59}\\
Leakage current &{\color{black}0.14}\\
ODT power effect &{\color{black}0.09}\\
E-squared effect &{\color{black}0.04}\\
Total &{\color{black}1.18} \\
\end{tabular}
\end{ruledtabular}
\end{table}

$\textit{EDM\ results\ -\ }$The systematic error budget for the EDM measurement is provided in Table~\ref{tab:table1} (more details are provided in the Supplement). {\color{black}We constantly monitor the current supplying the bias B field, and find the correlation between its fluctuation and the E field polarity to be less than the current-measurement uncertainty.} The residual parity-mixing  systematic  effect is determined by taking the difference of EDM measurements between ODT$_+$ and ODT$_-$. The effect comes from the residual misalignment of the polarization direction ($\pm6.7$ mrad) between ODT$_+$ and ODT$_-$, as well as the intensity imbalance ($\pm10\%$) between the two. The phase shifts as the E field is switched between {\color{black}73 kV/cm} and 0 kV/cm are measured to constrain the E-squared effects. The phase shifts as the ODT power is switched between {\color{black}14 W} and {\color{black}10 W} are measured to constrain the ODT power effect. Combining all the contributions in Table~\ref{tab:table1} and averaging the EDM results of ODT$_+$ and ODT$_-$, we arrive at the final result $d({\rm^{171}Yb})={\color{black}(-6.8\pm5.1_{\rm stat}\pm1.2_{\rm syst})\times10^{-27}\ e\ \rm cm}$, based on which we set a 95\% confidence limit $|d({\rm^{171}Yb})|<{\color{black}1.5\times10^{-26}\ e\ \rm cm}$. 

{\color{black} The atomic EDM result can be translated into an upper limit on the nuclear Schiff moment of $^{171}$Yb based on atomic structure calculations \cite{Flambaum20,Rad14,Sahoo17}, and is further related to constraints on BSM models according to particle and nuclear physics theories \cite{Dzuba07}. Compare with other diamagnetic systems that probe $P,T$-odd nuclear interactions (Table~\ref{tab:table2}), the constraints based on the $^{171}$Yb measurements of this work lag behind those of $^{199}$Hg, and are on the same order of magnitude as those of $^{129}$Xe, $^{225}$Ra, and $^{205}$Tl$^{19}$F. A global analysis of EDM results show that combining these different systems with complementary sensitivities to BSM parameters together can indeed improve the constraints set by $^{199}$Hg alone \cite{ChuRam15}.

}

\begin{table*}[htp]
\caption{\label{tab:table2}
{\color{black}Experimental upper limit of EDM $d_{\rm UL}$ ($95\%$ C.L.), upper limit of nuclear Schiff moment $S_{\rm UL}$, and theoretical calculation of Schiff moment $S_{\rm Th}$ in different isotopes. $\eta's$ are parameters of the BSM $P,T$-odd interactions.}}

\begin{ruledtabular}
\begin{tabular}{lccccc}
Diamagnetic system & $^{205}$Tl$^{19}$F & $^{199}$Hg & $^{129}$Xe & $^{225}$Ra & $^{171}$Yb \\

$d_{\rm UL}$/($10^{-26}\ e\ {\rm cm}$) &$6500$ \cite{ChuFieRam19,ChoSanHin91}&$7.4\times10^{-4}$ \cite{GraCheLin16,GraCheLin17}&0.14 \cite{SacFanBab19} &1400 \cite{BisParBai17} &1.5 [this work]\\

$S_{\rm UL}$/($e$ fm$^3$)$\times 10^{10}$ \footnotemark[1] &$8.8$ \cite{ChuFieRam19,Flambaum20} &$2.6\times10^{-3}$ \cite{ChuFieRam19,Flambaum20}&5.2 \cite{ChuFieRam19,Flambaum20}&1700 \cite{ChuFieRam19,Flambaum20}& $7.9$ \cite{Flambaum20,Rad14,Sahoo17} \\

$S_{\rm Th}$/($e$ fm$^3$)$\times 10^8$ &$1.2\ \eta_{\rm pp}-1.4\ \eta_{\rm pn}$ \cite{Flambaum86}&$-1.4$ $\eta_{\rm np}$ \cite{Flambaum86} &1.75 $\eta_{\rm np}$ \cite{Flambaum86} &300 $\eta_{\rm n}$ \cite{FlaZel03}, 1100 $\eta$ \cite{AueFlaSpe96} &$\sim-1.4$ $\eta_{\rm np}$ \cite{Dzuba07}\\

\end{tabular}
\end{ruledtabular}
\footnotetext[1]{Calculated based on the conversion factors between atomic (molecular) EDM and nuclear Schiff moment given in \cite{ChuFieRam19,Flambaum20,Rad14,Sahoo17}}
\end{table*}

$\textit{Outlook\ -\ }$In the next phase of the experiment aiming for a higher EDM sensitivity, magnetometers or a co-magnetometer need to be implemented to alleviate the B-field noise problem and the B-field-induced systematic effect. In order to further suppress the parity-mixing  systematic  effect, atoms can be held in an optical lattice inside an optical cavity, an arrangement that can guarantee a much better intensity balance between the $+k$ and $-k$ optical fields. {\color{black}In addition, electrodes capable of generating $E=500\ \rm kV/cm$ has recently been demonstrated \cite{RoyGorKev21}, and a spin precession time $\tau=300\ \rm s$ can be realized with an improved vacuum. With an atom number $N=1\times 10^6$, and an integration time of $T=100$ days, the ODT method explored in this work could reach an EDM sensitivity of $2\times 10^{-29}\ e\ \rm cm$.} 

The cold-atom and EDM measurement techniques realized in this work, particularly the highly efficient QND detection scheme and various ways of controlling systematic effects, can be transferred to the study of $^{225}$Ra EDM. Furthermore, $^{171}$Yb can itself act as a co-magnetometer in a $^{225}$Ra EDM measurement. {\color{black}Due to nuclear-octupole enhancement, a $^{225}$Ra EDM measurement at $10^{-27}\ e\ \rm cm$ would directly rival that of $^{199}$Hg in the search for BSM physics.}

\begin{acknowledgments}
We would like to thank K. G. Bailey, M. Bishof, C.-F. Cheng, M. R. Dietrich, W.-K. Hu, M.-D. Li, Y.-N. Lv, P. Mueller, T. P. O'Connor, D. Sheng, Z.-S. Yuan, C.-L. Zou, Y.-G. Zheng, W. Jiang for helpful discussions. This work has been supported by the National Natural Science Foundation of China through Grants No. 91636215, No. 12174371, No. 11704368, the Strategic Priority Research Program of the Chinese Academy of Sciences through Grant No. XDB21010200, and Anhui Initiative in Quantum Information Technologies through Grant No. AHY110000.
\end{acknowledgments}


\nocite{*}



%

\end{document}


\title{Supplemental Material for “Measurement of the Electric Dipole Moment of $^{171}$Yb Atoms in an Optical Dipole Trap”}

\author{T. A. Zheng$^{1}$}
\thanks{These authors contributed equally to this work.}
\author{Y. A. Yang$^{1}$}%
\thanks{These authors contributed equally to this work.}
\author{S.-Z. Wang$^{1}$}
\author{J. T. Singh$^{2}$}
\author{Z.-X. Xiong$^{3}$}%
\author{T. Xia$^{1}$}%
\email{txia1@ustc.edu.cn}
\author{Z.-T. Lu$^{1}$}%
\email{ztlu@ustc.edu.cn}

\affiliation{$^1$CAS Center for Excellence in Quantum Information and Quantum Physics, University of Science and Technology of China, Hefei 230026, China}%
\affiliation{$^2$National Superconducting Cyclotron Laboratory and Department of Physics and Astronomy,
Michigan State University, East Lansing, Michigan 48824, USA}%
\affiliation{$^3$Key Laboratory of Atomic Frequency Standards, Innovation Academy for Precision Measurement Science and Technology, Chinese Academy of Sciences, Wuhan 430071, China}


\maketitle

\section*{systematic effects} \label{sec:systematic effect}
\subsection{Bias B-field correlation}
Any change of the B field during spin precession that correlates with the polarity of the static E field produces a false EDM signal
\begin{equation}
d_{\rm false}=\frac{\mu\Delta B}{2E},
\label{B_field_corre}
\end{equation}
where $\Delta B$ is the difference in B field between the two precession periods. The B field is proportional to the current that supplies the field coil. Throughout the EDM measurements, the current is constantly monitored by a 6-digit multimeter (Keysight 34465A). The measured change of current corresponds to $\Delta B=(0.9\pm9.5)\times10^{-3}$ pT . The corresponding false EDM is $d_{\rm false}=(0.1\pm1.0)\times10^{-27}\ e\ \rm cm$. The contribution to the EDM systematic uncertainty is set to be $1.0\times10^{-27}\ e\ \rm cm$. This systematic is likely to reduce further as the precision of current measurement improves. 

\subsection{Residual parity-mixing systematic effect}
The residual parity-mixing systematic effect is given by 
\begin{equation}
d_{res}=\pm\ c_p\ I_{+/-}\ \sin(2\theta_{+/-}),
\label{stark_res}
\end{equation}
where $c_p$ is the coefficient for the parity-mixing systematic effect, $I_{+/-}$ is the intensity for ODT$_{+/-}$, and  $\theta_{+/-}\approx\pi/2$ is the polarization angle of the ODT$_{+/-}$. The fitting of the parity-mixing effect measurement (Fig. 4 in main text) gives the amplitude
\begin{equation}
c_p\ I_p=-10.6\ (1.7)\times10^{-26}\ e\ {\rm cm}.
\label{amp}
\end{equation}
Where $I_p$ is the ODT intensity used to measure the parity-mixing effect, and is 3.2(3) times that of $I_+$. The relation between intensities of ODT$_{+/-}$ is determined by measuring the beam waists and powers of the ODT$_{+/-}$, 
\begin{equation}
f=I_{-}/I_{+}=1.02\,(0.10).
\label{f}
\end{equation}

The EDM results obtained in ODT$_+$ and ODT$_-$ are:
\begin{equation}
d_+={d}-2\ c_p\  I_{+}\ \Delta \theta_{+}
\label{d+}
\end{equation}
\begin{equation}
d_-={d}+2\ c_p\ I_{-}\ \Delta \theta_{-}
\label{d-}
\end{equation}
where $d$ is the true EDM and $\Delta \theta_{+/-}=\theta_{+/-}-\pi/2\ll1$.

The difference between the EDM results is used to determine the value of $\Delta \theta_{+/-}$,
\begin{equation}
d_+-d_-=-2\ c_p\ I_{+}\ (\ \Delta\theta_{+}+f\ \Delta\theta_{-}).
\label{d+-d-}
\end{equation}
The difference of polarization angles between ODT$_+$ and ODT$_-$ is measured to a precision of 6.7 mrad,
\begin{equation}
\Delta\theta_{-}=\Delta\theta_{+}\pm0.0067.
\label{angle_diff}
\end{equation}
The two measurement results are
\begin{equation}
d_{+}=(0.41\pm0.82)\times10^{-26}\ e\ \rm cm,
\label{d+EDMvalue}
\end{equation}
and
\begin{equation}
d_{-}=(-1.77\pm0.61)\times10^{-26}\ e\ \rm cm.
\label{d-EDMvalue}
\end{equation}
By substituting Eq. (\ref{amp}), Eq. (\ref{angle_diff}), Eq. (\ref{d+EDMvalue}) and Eq. (\ref{d-EDMvalue}) into Eq. (\ref{d+-d-}), we get
\begin{equation}
\Delta\theta_{+}=+0.16\pm0.08.
\label{dtheta+value}
\end{equation}

Take the average of the two EDM results, we get

\begin{eqnarray}
{d}&=&\frac{{ d_+}+{ d_-}+2\ c_p\ I_+\ (\Delta \theta_+-f\ \Delta \theta_-)}{2}\\
&=&\frac{{ d_+}+{ d_-}+2\ c_p\ I_+\ [(1-f)\ \Delta \theta_+ + f\ (\Delta\theta_+-\Delta\theta_-)]}{2}.
\label{EDM_1}
\end{eqnarray}

With the results of Eq. (\ref{amp}), Eq. (\ref{f}), Eq. (\ref{angle_diff}), and Eq. (\ref{dtheta+value}), the residue parity mixing effect is determined to be $d_{false}=\pm5.9\times10^{-28}\ e\ {\rm cm}$. In order to further suppress the parity-mixing systematic effect, atoms can be held in an optical lattice inside an optical cavity, an arrangement that can guarantee a much better intensity balance between the $+k$ and $-k$ optical fields, further reducing the value of $|1-f|$.

\subsection{E-squared effect}
Although the static E field and B field are nearly uniform, the small E field gradient can shift the position of the atoms to where the B field is slightly different due a B field gradient. The effect can be eliminated if the magnitude of the E field is identical upon E field reversal. The E-squared effect is the residue of this high-order effect due to imperfections in the E-field uniformity, the B-field uniformity, and the E-field reversal. To evaluate this systematic effect, we measure the phase shift between the E field at 73 kV/cm and 0 kV/cm. The measured phase shift difference between E-field on and E-field off is equivalent to a false EDM of $(1.9\pm0.1)\times10^{-24}\ e\ \rm cm$. During the EDM data run, the E field is constantly monitored by a calibrated high-voltage divider, and the E-field imbalance between the two polarities is measured to be less than 1$\times10^{-5}$. The contribution to the uncertainty of the EDM measurement is then $4\times10^{-29}\ e\ \rm cm$.

\subsection{ODT power effect}
Phase shifts can be induced by changes of the ODT power. Possible causes are: 1) residue vector shifts induced by the ODT; 2) shifts of the trap position in the ODT coupled with B-field gradients. A correlation between the ODT power and the E-field polarities, which could arise from a ground connection between the power-stabilization module of the ODT and the high voltage system. This effect could then cause a systematic effect in the EDM measurements. To evaluate this effect, we measure the phase shift between the ODT power of 14 W and 10 W, and find the measured phase shift equivalent to an EDM of $(-4.1\pm0.2)\times10^{-24}\ e\ \rm cm$. During the EDM data run, the ODT power is constantly monitored. The ODT power difference between the two polarities is measured to be less than 2$\times10^{-5}$. The contribution to the uncertainty of the EDM measurement is then $9\times10^{-29}\ e\ \rm cm$.

\subsection{Leakage current}
A leakage current conducted between the electrodes generates a magnetic field, and is naturally correlated with the E field reversal. When the current, assumed to be in the direction of the E field, is not exactly aligned with the static B field, this effect generates a false EDM \cite{BisParBai17}:
\begin{equation}
{d_{\rm false}}=\frac{\mu}{E}\frac{\mu_0 I}{2\pi r}\sin{\theta_{\rm EB}},
\label{LeakageC}
\end{equation}
where $\mu_0$ is the vacuum permeability, $I$ is the leakage current, $r$ is the distance between the atoms and the current, and $\theta_{\rm EB}$ is the misalignment angle between the static E field and B field. We consider the worst case, where the current is along a path close to the atoms, $r=50\ \mu$m, which is the ODT beam waist. We use a conservative upper limit for the field misalignment, $\theta_{\rm EB}\leq0.1$ rad, in our experiment. The leakage current measured during the EDM data run is $-1.60(2)$ pA. With Eq. (\ref{LeakageC}), we get the systematic error due to the leakage current $d_{false}=1.4\times10^{-28}\ e\ {\rm cm}$.

\nocite{*}

%